\documentclass[
showpacs,preprintnumbers,amsmath,amssymb,pre]{revtex4}

\usepackage{graphicx}
\usepackage{dcolumn}
\usepackage{bm}
\font\msbmtex=msbm10

\def\RSp{\mbox{{\msbmtex R}}}
\def\ZSp{\mbox{{\msbmtex Z}}}
\begin{document}

\preprint{}

\title{Wavelet treatment of the intra--chain correlation\\ functions
of homopolymers in dilute solutions}

\author{M.V.~Fedorov}
\email{Maxim.Fedorov@ucd.ie}
\affiliation{%
Theory and Computation Group, Centre for Synthesis and Chemical
Biology, Conway Institute of Biomolecular and Biomedical Research,
Department of Chemistry, University College Dublin, Belfield,
Dublin 4, Ireland
}%

\author{G.N.~Chuev}
\email{genchuev@rambler.ru} \affiliation{ Institute of Theoretical
and Experimental Biophysics, Russian Academy of Sciences,
Pushchino, Moscow Region, 142290, Russia
}%

\author{Yu.A.~Kuznetsov}
\email{Yuri.Kuznetsov@ucd.ie} \affiliation{ Computing Centre,
University College Dublin, Belfield, Dublin 4, Ireland
}%

\author{E.G.~Timoshenko}
\homepage{http://darkstar.ucd.ie} \email{Edward.Timoshenko@ucd.ie}
\thanks{Author to whom correspondence should be addressed.}
\affiliation{ Theory and Computation Group, Centre for Synthesis
and Chemical Biology, Conway Institute of Biomolecular and
Biomedical Research, Department of Chemistry, University College
Dublin, Belfield, Dublin 4, Ireland
}%

\date{\today}

\begin{abstract}
Discrete wavelets are applied to parametrization of the intra--chain
two--point correlation functions of homopolymers in
dilute solutions obtained from Monte Carlo
simulation. Several orthogonal and biorthogonal basis sets have been
investigated for use in the truncated wavelet approximation.
Quality of the approximation has been assessed by calculation
of the scaling exponents obtained from des Cloizeaux ansatz for
the correlation functions of homopolymers with different
connectivities in a good solvent. The resulting exponents are
in a better agreement with those from the recent renormalisation
group calculations as compared to the data without the wavelet
denoising. We also discuss how the wavelet treatment improves the
quality of data for correlation functions from simulations of
homopolymers at varied solvent conditions and of heteropolymers.
\end{abstract}


\maketitle

\section{\label{sec:intoroduction}Introduction}

The main purpose of this paper is to give a useful introduction
and a practical guide to those who would like to apply discrete
wavelets for treating the data for the intra--chain two--point
correlation functions (TPCF) $g_{ij}^{(2)}(r)$, which either have
been previously computed from direct computer simulations, came
from some theoretical technique after solving equations for TPCFs,
or perhaps have been obtained from X--ray and neutron scattering
experiments. The intra--chain correlation functions represent a
fundamental link between the equilibrium thermodynamic observables
and the conformational structure of polymers. These functions for
polymers exhibit rather different behavior depending on the
solvent quality. TPCF of a homopolymer in a good solvent follows a
universal scaling scaling law 
\footnote{Strictly speaking \cite{IntraCor}, such laws are
asymptotic in nature and do not apply when the two monomers are
too close to each other in terms of the connectivity, or when the
interaction parameters are far from the the appropriate fixed
point.}
for which analytical
expressions can be derived by the field theoretical and other
approaches \cite{CloizeauxBook,theory,Schafer}. On the contrary,
the TPCF in a poor solvent exhibit a complicated oscillating
radial dependence akin to that of simple liquids. In this case,
there is no known simply parametrized representation of TPCF for
the homopolymer globule. Moreover, an accurate sampling around a
rather tall peak corresponding to the first solvation shell
becomes very significant as this peak contributes most to the
thermodynamic observables such as the mean energy. On the other
hand, TPCF in a good solvent obtained from molecular mechanics
simulations tend to be rather noisy due to the high entropy of the
coil conformation. This results in a large scatter of values of
TPCF at small radial separations, which makes further fitting of
the data by an analytical expression and extraction of the scaling
exponents difficult. Therefore, in general, dealing with TPCF data
of heteropolymers, for which some monomers are in a good solvent
while others are in a poor solvent, and, particularly, extracting
meaningful information from such data is a rather nontrivial
problem.

Relying on the recent works of some of us
\cite{chuevfedorov1,chuevfedorov2,chuevfedorov3},
we believe that the task of parameterizing $g_{ij}^{(2)}(r)$
in a compact way can be accomplished by means of the
multiresolution analysis \cite{mallat,meyer}.
At present, a number of special  basis sets, referred to as wavelets
\cite{daubechies}, are known and are being actively used for treating
both smooth  and sharply oscillating functions, as well as for denoising
of signals \cite{donoho,ogden}.
Wavelets became a necessary mathematical tool in many modern theoretical
investigations in Physics, Chemistry and other fields
\cite{cho,wei,goedecker1,goedecker2,han,ivanov,antoine,arias,kolb1,kolb2,romeo}.
Wavelets are particularly useful in those cases when the result
of the analysis of a function should contain not only the list of its
typical frequencies (scales), but also the list of the
local coordinates where these frequencies are important.
Thus, the main field of applications of wavelets is to analyse and
process different classes of functions which are either nonstationary (in
time) or inhomogeneous (in space).

The most general principle of the wavelet construction is to use
dilations and translations.
Commonly used wavelets form a complete (bi)orthonormal system of
functions with a finite support constructed in such a way. That is
why by changing a scale (dilations) wavelets can distinguish the local
characteristics of a function at various scales, and by
translations they cover the whole region in which a function is
being studied.
Due to the completeness of the base system, wavelets also allow one
to perform the inverse transformation to decomposition, which is
called reconstruction.

In the analysis of functions with a complicated behavior, the locality property
of wavelets makes the wavelet transform technique substantially advantageous
compared to the Fourier transform. The latter provides
one only with the knowledge of global frequencies (scales) of a
function under investigation since the system of the base
functions used (sine, cosine or imaginary exponential functions)
is defined on the infinite range. The special features of
wavelets such as their (bi)orthogonality and vanishing of moments result in
the need for only few approximating coefficients in practical applications.
That is a reason why wavelets are actively used, for example,
to construct distribution functions in calculations of the electronic
structure \cite{arias,kolb1,kolb2}
as well as in Statistical Mechanics
\cite{chuevfedorov1,chuevfedorov2,chuevfedorov3}.

Recently, some of us, have carried out several studies devoted
to the wavelet parametrization of the radial density functions for
various atomic and molecular solutes
\cite{chuevfedorov1,chuevfedorov2,chuevfedorov3}. A model study
of the galaxies density in Ref.~\cite{romeo} uses a similar
wavelet approach for a different problem. In the present work
we would like to address the question whether wavelets can also be
advantageous for approximating the intra--chain correlation
functions of homopolymers in different solvents.
The main practical goal of this paper is to apply discrete
wavelets for approximating functions $g_{ij}^{(2)}(r)$ of open, ring
and star homopolymers in a coil conformation, as well as of a globule.
In the case of a coil, the des Cloizeaux scaling formula applies and a number
of accurate theoretical results for the scaling exponents involved are
available  \cite{CloizeauxBook,Guida}. Thus, we shall be able to investigate
the influence of the choice of the wavelet basis set and of the number of terms
not only on the quality of the correlation function parameterization,
but also on the values of the scaling exponents extracted from fitting the
wavelet denoised functions by the des Cloizeaux formula.

\section{Methods}

\subsection{Model}

To obtain the correlation functions we
relied on the standard coarse--grained homopolymer model
\cite{CombStar,Torus,CopStar} based on the following Hamiltonian
in terms of the monomer coordinates, ${\bf X}_i$:
\begin{eqnarray}
H & = & \frac{k_B T}{2\ell^2}  \sum_{i\sim j} \kappa_{ij}
        ({\bf X}_i - {\bf X}_{j})^2 \nonumber\\
  &+& \frac{1}{2} \sum_{ij,\ i\not= j} V (|{\bf X}_i - {\bf X}_j|).
\label{cmc:hamil}
\end{eqnarray}
The first term here represents the connectivity structure of the polymer with
harmonic springs of a given strength $\kappa_{ij}$ introduced between
any pair of connected monomers (denoted by $i\sim j$).
The second term represents pair--wise non--bonded interactions
between monomers such as the van der Waals forces, for which
we adopt the Lennard--Jones form of the potential,
\begin{equation}\label{VLJ}
V(r) = \left\{
\begin{array}{ll}
+\infty, &  r < d \\
V_0 \left( \left( \frac{d}{r}\right)^{12}
- \left( \frac{d}{r} \right)^{6} \right), &  r > d
\end{array}
\right.,
\end{equation}
where there is also a hard core part with the monomer diameter $d$
(below we choose $d=\ell$ without any lack of generality).

We use the Monte Carlo technique with the standard Metropolis
algorithm \cite{AllenTild}, which converges to the Gibbs
equilibrium ensemble, based upon the implementation described by
us in \cite{Torus}. Value of $V_0=0$ will correspond to
the purely repulsive case (good solvent) leading to a coil
conformation of the polymer, while $V_0=5\,k_B T$
will correspond to the attractive case (poor solvent)
leading to a globular conformation as in Ref.
\cite{IntraCor}.
All details of our Monte Carlo procedure have been previously described
in the paper \cite{IntraCor} and, in fact, here we shall rely on the same
set of Monte Carlo simulation data in order to make the comparison of the
wavelet treated scaling exponents with those of Ref. \cite{IntraCor} more
straightforward and unambiguous.

\subsection{Correlation functions}

The intra--chain two--point correlation function (TPCF) of
a pair of monomers $i$ and $j$ is defined as,
\begin{equation}
g^{(2)}_{ij}({\bf r}) \equiv \bigl\langle \delta({\bf X}_i
- {\bf X}_{j}-{\bf r}) \bigr\rangle = \frac{1}{4\pi r^2} \bigl\langle
\delta(|{\bf X}_i - {\bf X}_{j}|-r) \bigr\rangle.
\end{equation}
The second equation establishes that it is a function of radius
$r=|{\bf r}|$ only due to spatial isotropy (SO(3) rotational
symmetry). We may note that this function should, strictly
speaking, be named distribution function, but since
$g^{(2)}_{ij}(r)\to 0$ when $r\to \infty$ because of the chain
connectivity, we apply the term `correlation function'  to
$g^{(2)}_{ij}(r)$ itself rather than to the quantity
$g^{(2)}_{ij}(r)/(g^{(1)})^2-1$, which would vanish as $r\to
\infty$  in the case of simple liquids. The function is normalized
to unity via: $\int d^3{\bf r}\, g^{(2)}_{ij}({\bf r})=1$. Note
that the correlation functions exactly satisfy the excluded volume
condition, $g_{ij}^{(2)}(r)=0$ for $r < d,$
due to the choice of the hard--core part in the non--bonded potential
Eq. (\ref{VLJ}).
The mean--squared distance between monomers $i$ and $j$ is,
\begin{equation}\label{Dd}
D_{ij} \equiv   \biggl\langle ({\bf
X}_i - {\bf X}_{j})^2 \biggr\rangle = \int d^3{\bf r}\,|{\bf
r}|^2\, g^{(2)}_{ij}({\bf r}),
\end{equation}
which we defined here without the traditional factor of $1/3$ as compared
to some of the previous papers \cite{Torus}.

The intra--chain pair correlation
functions $g^{(2)}_{ij}(r)$ are strongly dependent on both
the degree of polymerization $K$ of the polymer and the choice of
the reference monomers $i$
and $j$, contacts between which we are looking at. However, as we
have demonstrated in Ref. \cite{IntraCor}, if we
introduce the rescaled correlation function in terms of the
dimensionless variables,
\begin{equation}
\hat{g}^{(2)}_{ij}(\hat{r}) \equiv { D}_{ij}^{3/2}\,\,
g^{(2)}_{ij}\left(r\right), \qquad \hat{r} \equiv r/{
D}_{ij}^{1/2},
\end{equation}
these will change in about the same range and hence would
permit a much more straightforward comparison with each
other. From this definition, obviously, $\hat{g}^{(2)}_{ij}(\hat{r})$
satisfies the following two normalization conditions:
\begin{equation}
\label{normtwo}
\int_0^{\infty} d\hat{r}\,\hat{r}^2\,\hat{g}^{(2)}_{ij}(\hat{r})=
\int_0^{\infty} d\hat{r}\,\hat{r}^4\,\hat{g}^{(2)}_{ij}(\hat{r})= \frac{1}{4\pi}.
\end{equation}

\subsection{Scaling Relations}

According to Refs. \cite{CloizeauxBook,Schafer} TPCF of a
flexible homopolymer coil in a good solvent can be well described
\cite{IntraCor} via
a power law times a stretched exponential, known as the des Cloizeaux
scaling equation,
\begin{equation} \label{FitA}
\hat{g}^{(2)}_{ij}(\hat{r}) = A_{ij}\,\hat{r}^{\theta_{ij}}\,
\exp\left( -B_{ij}\,\hat{r}^{\delta_{ij}} \right).
\end{equation}
Due to the
two normalization conditions in Eq. (\ref{normtwo}) constants $A$
and $B$ can be immediately calculated and expressed via $\theta$
and $\delta$.
The exponents $\delta_{ij}$ do not really depend on $i,j$,
but the contact exponents $\theta_{ij}$ do. In the case of the end--end
correlations of an open chain $\theta_{ij}$ is denoted as $\theta_0$,
and these can be expressed via,
\begin{equation}\label{DeCl}
\delta=\frac{1}{1-\nu}, \qquad \theta_0=\frac{\gamma-1}{\nu},
\end{equation}
where $\nu$ has the meaning of the inverse fractal dimension of the system
and $\gamma$ is related to the number of different polymer conformations
\cite{CloizeauxBook, Schafer}.

\subsection{Wavelet Theory}

The fundamental theory behind wavelets is known as the
Multi--Resolution Analysis (MRA). Most of the rigorous results and
definitions from MRA are not usually required for practical
applications. The only equations which are needed for the work
described herein will be introduced in this section. As we mainly
use basis sets from the biorthogonal wavelets families, we shall
introduce all wavelets in a general way as biorthogonal wavelets.
Moreover, we shall use the Discrete Wavelet Transform (DWT)
technique \cite{mallat,daubechies} to parameterize the TPCFs.
There is a good introduction to the wavelet techniques in
Ref.~\cite{goedecker2}. We also will follow the style of that book
henceforth. The multiresolution  approach is based on the idea that the wavelet
functions generate   hierarchical sequence of
subspaces in the space of square--integrable
functions over the real axis $L^{2}(\RSp)$, which forms the MRA.

The scaling functions $\varphi (r)$ and $\tilde{\varphi}
(r)$  produce a biorthogonal MRA if they satisfy the following
conditions.

(i) Translates of these functions with integers $\varphi _{s}=\varphi (r-s)$,
$\tilde{\varphi}_{s}=\tilde{\varphi} (r-s)$, $ s \in {\ZSp}$, are
linearly independent and produce  bases of the subspace
$V_{0}\subset L^{2}(\RSp)$ and their dual counterpart
$\tilde{V}_{0}\subset L^{2}(\RSp)$ correspondingly. This means that
if a function $f(r)$ is contained in the space $V_{j}$, its
integer translates have to be contained in the same space,
\begin{equation}
f(r) \in V_{j} \Leftrightarrow f(r+s) \in V_{j}, \qquad  f(r) \in \tilde{V}_{j}
 \Leftrightarrow f(r+s) \in \tilde{V}_{j}, \quad s \in {\ZSp}. \notag
 \end{equation}

(ii) Dyadic dilates of these functions $ \varphi _{js}=\varphi
(2^{j}r-s)$, $ \tilde{\varphi} _{js}=\tilde{\varphi} (2^{j}r-s)$,
 $j\in {\ZSp}$, generate hierarchical sets of subspaces $\{V_ {j}\}$ and
$\{\tilde{V} _{j}\}$ , so that:
\begin{eqnarray}
&& V_{j}\subset V_{j+1}
,\qquad \bigcup\limits_{j=-\infty }^{\infty }V_{j}\; \mbox{is dense in} \;
L^{2}(\mathbb{R}),\qquad \bigcap\limits_{j=-\infty }^{\infty }V_{j}=0,  \label{Vsequence}\\
&& \tilde{ V} _{j}\subset \tilde{  V} _{j+1} ,\qquad
\bigcup\limits_{j=-\infty }^{\infty }\tilde{ V}_{j}
\; \mbox{is dense in} \;
L^{2}(\RSp),\qquad \bigcap\limits_{j=-\infty }^{\infty }\tilde{
V}_{j}=0. \notag
\end{eqnarray}

(iii) The  sets of  functions $\varphi_{js}(r)$ and
$\tilde{\varphi}_{js}(r)$ are biorthogonal to each other. It
means that for  any $s, s' \in {\ZSp}$:
\begin{equation}
\int \varphi _{js}(r)\tilde{\varphi}_{js'}(r)\,dr=
\delta _{ss'}.  \notag
\end{equation}
 It means that if a function $f(r)$ is contained in the space $V_{j}$,
 the compressed function $f(2r)$ has to be contained in the higher resolution space $V_{j+1}$
 \begin{equation}
 f(r) \in V_{j} \Leftrightarrow f(2r) \in V_{j+1}, \qquad  f(r) \in \tilde{V}_{j}
 \Leftrightarrow f(2r) \in \tilde{V}_{j+1}, \quad j \in {\ZSp}. \notag
 \end{equation}

(iv) There is a wavelet function $\psi(r)$ and its dual wavelet
function $\tilde{\psi}(r)$ such that their integer translates $
\psi_{s}(r)=\psi(r-s)$, $\tilde{\psi_{s}}(r)=\tilde{\psi}(r-s)$,
and  dyadic dilates $ \psi _{js}=\psi (2^{j}r-s)$, $\tilde{\psi}
_{js}=\tilde{\psi} (2^{j}r-s)$,  form subspaces $W_{j}$ and
$\tilde{W_{j}}$ which are complementary to $V_{j}$ and
$\tilde{V_{j}}$ so that:
\begin{equation}
V_{j+1}=V_{j}\oplus W_{j}, \qquad \tilde{ V}_{j+1}=\tilde{V}
_{j}\oplus \tilde{W} _{j}, \qquad
 \tilde{W} _{j} \bot V_{j}, \qquad \tilde{V} _{j} \bot  W_{j}.
\end{equation}

(v) From the above relations it follows that $L^{2}(\RSp)$ can be
decomposed into the approximation space $V_{j_{0}^{{}}}$ and the sum
of the detailed spaces $W_{j}$ of higher resolutions $j \geqslant
j_{0}$:
\begin{equation}
L^{2}(\RSp)=V_{j_{0}}\oplus \bigoplus_{j\geq j_{0}}^{\infty }W_{j}, \label{SpaceDecomposition}
\end{equation}
where $j_{0}\in {\ZSp}$ is a chosen level of resolution. This means
that any square--integrable function $f(r)$ can be represented as a
sum of linear combinations of the reconstruction scaling functions
$\{\varphi_{j_{0}}\}$ at a chosen resolution $j=j_{0}$ and the
reconstruction wavelet functions $\{\psi_{j}\}$ at all finer
resolutions $j\geq j_{0}$. This can be written as,
\begin{equation}
f(r)=\sum_{s}a_{j_{0}s}\varphi_{j_{0}s}(r)+\sum_{j>j_{0}}^{\infty
}\sum_{s}d_{js}\psi _{js}(r),  \label{WaveReprFunction}
\end{equation}
where the coefficients $\{a_{j_{0}s}\}$ and $\{d_{js}\}$ are obtained as the
scalar products with the appropriate dual decomposition basis functions,
\begin{equation}
a_{js}=\int f(r)\tilde{\varphi}_{js}(r)\,dr,\quad
d_{js}=\int f(r)\tilde{\psi}_{js}(r)\,dr.
\label{WaveConvolution}
\end{equation}
The later equation defines the Discrete Wavelet Transform (DWT).

As $\varphi (r)\subset V_{0}$  and $ V_{0}\subset V_{1}$,
$\tilde{\varphi} (r)\subset\tilde{ V}_{0}$  and
$\tilde{V}_{0}\subset \tilde{V}_{1}$,
we can express $%
\varphi (r)$ (as well as $\tilde{\varphi}(r)$) as a linear combination of the  basis functions in $V_{1}$ ($\tilde{V}_{1}$):
\begin{equation}
\varphi (r)=\sum\limits_{s}h_{s}\varphi (2r-s), \qquad
\tilde{\varphi} (r)=\sum\limits_{s}\tilde{h}_{s} \tilde{\varphi}
(2r-s).  \label{DilationPhi}
\end{equation}
This equation is called  the  {\it dilation equation}. Similarly, $\psi (r)$ and $\tilde{\psi} (r)$
must satisfy a {\it wavelet dilation equation}:
\begin{equation}
\psi (r)=\sum\limits_{s}w_{s}\varphi (2r-s), \qquad \tilde{\psi}
(r)=\sum\limits_{s} \tilde{w}_{s} \tilde{\varphi} (2r-s).
\label{DilationPsi}
\end{equation}
The above sets of coefficients are usually called "filters" and they
are completely sufficient in order to describe a chosen wavelet basis because
there are several procedures on how to build up numerical values of
the wavelet functions from the set of filters
\cite{daubechies,mallat,goedecker2}. We should emphasize here that
there are no analytic expressions for biorthogonal
(orthogonal) wavelets with a finite support
\footnote{This is true except of the simplest basis, Haar basis, which
is constructed from piece--wise functions \cite{daubechies}.}.
These are
determined in terms of their filter coefficients only. But one can obtain
the values of these functions with any given accuracy by using special
procedures, which are well described in the wavelet literature
\cite{daubechies,mallat,goedecker2}.

The scaling functions and the wavelets have a finite support only in the
case of a finite number of the coefficients $h_{s}$ and $w_{s}$. Due to their biorthogonal nature these
 functions satisfy the relations:
\begin{eqnarray}
\int \varphi _{ja}(r)\tilde{\varphi}_{jb}(r)\,dr=
\delta _{ab}, &&  \notag
\\
\int \varphi _{ja}(r)\tilde{\psi}_{lb}(r)\,dr=0,
\quad (l\geq j), &&  \label{WProperties}
\\
\int \tilde{\varphi}_{ja}(r)\psi _{lb}(r)\,dr=0,
\quad (j\geq l), &&  \notag
\\
\int\psi _{ja}(r)\tilde{\psi}_{kb}(r)\,dr=\delta _{jk}\delta _{ab}, &&  \notag
\end{eqnarray}
for any integer $j,l,a, b$.

If the pairs of the decomposition functions
$\{\tilde{\varphi},\tilde{\psi}\}$ and the reconstruction
functions $\{\varphi,\psi\}$ are identical, the transform is
called `orthogonal wavelet transform.' Otherwise we shall talk
about a more general `biorthogonal wavelet transform.'

In the expansion (\ref{WaveReprFunction}) the first term gives a
`coarse' approximation for $f(r)$ at the resolution $j_{0}$ and
the second term gives a sequence of successive 'details'. In
practice, we actually do not need to use the infinite number of
resolutions. Therefore, the sequence of details is cut--off at an
appropriate resolution $j_{\max }$. Since all functions used in
numerical work are given in a finite interval, the sequence of
different translates $\{s\}$ has also a finite number of terms
$S$. It should be mentioned that, really, $S$ can be different for
detailed and coarse approximations.

Importantly, the explicit form of the basis functions is not required if
we are using (bi)orthogonal wavelets with a finite support and a dyadic set of
scales {j}.  Then the coefficients in Eq.~(\ref{WaveConvolution})
can be calculated by the Fast Wavelet Transform (FWT) algorithm \cite{mallat,meyer,goedecker2}.
The main idea of this algorithm is that a set
of (bi)orthogonal discrete filters at consequently dilated scales
is used for the multi--resolution analysis of a signal. As a result,
to calculate the approximating coefficients, the convolution of the
signal and the relevant filter is only required for each scale, and
the latter can be easily obtained.

By choosing relevant basis functions and scales we can nullify most
of the coefficients $\{a\}$ and $\{d\}$ thereby reducing the
square root error (SRE)
since DWT satisfies the Parseval's identity
\cite{daubechies}. Therefore, the function under study can be reconstructed
with the use of only a few nonzero coefficients without any significant loss
of accuracy, making the total number of the approximating coefficients
rather small.
This feature of the of wavelet approximation is widely used
in processing of signals and images, the data for which should be
compressed with minimal losses \cite{donoho}.

\subsection{Choice of wavelet basis set.}

The compression and denoising  properties of the wavelet transform strongly depended on the 
fundamental properties of the wavelet bases, which we define here in a rather simplified way as:
{\it the number of vanishing moments, regularity, size of support,  symmetry and orthogonality/biorthogonality.}
\begin{description}
\item[Number of vanishing moments:] A wavelet function $\psi (r)$ has $N_{v.m.}$ vanishing moments if:
\begin{equation}
\int r^{\mu}\psi (r) dr=0 \quad \ \mbox{for} \quad \mu=0,\ldots , N_ {v.m.}-1, \label{VanishingMoments}
\end{equation}
The number of vanishing moments strongly influences the localization of wavelets in the frequency space.
The Fourier transform of a wavelet with $N_ {v.m.}=n$ has a peak and decays
as $k^{-n}$ ($k$ means frequency).
\item[Regularity:] This can be defined as the number $\rho$ of existing derivatives of a wavelet function.  It also characterizes the frequency localization 
of wavelets. The Fourier transform of a wavelet with regularity $\rho=n$ decays as $k^{-(n+1)}$ for large $k$.
We would like to emphasize that as wavelets have no analytic expressions the definition of their derivatives is not
as straightforward as for "usual" functions \cite{daubechies}. However, 
these mathematical details are beyond of our article.
\item[Size of support:] This is the length of the interval on which the  wavelet function has non-zero values. Obviously, this characterizes
the space localization of the wavelet.
\item[Symmetry:] The wavelet bases functions can be strongly symmetric or asymmetric.
The deviation of a  wavelet from the symmetry (i.e. even or odd parity) is usually measured by how
the phase of its Fourier transform deviates from a linear function.
   It was shown that is impossible to construct an orthogonal basis
 with the exact parity of the functions 
\footnote{The Haar basis is also an exception in this case.}.

 On the contrary, we can design a
 biorthogonal basis set with the exact symmetry of the
 function without serious
 efforts \cite{daubechies,sweldens}.
\item[Orthogonality/Biorthogonality:] As we have already mentioned in the case
when the pairs of the decomposition functions
$\{\tilde{\varphi},\tilde{\psi}\}$ and the
reconstruction functions $\{\varphi,\psi\}$
are identical, the wavelet transform is orthogonal.
Otherwise it is
biorthogonal. But this is true only  if the $\{\tilde{\varphi},\tilde{\psi}\}$ and $\{\varphi,\psi\}$
obey the conditions (\ref{WProperties}). We should mention that there are several non-orthogonal families of wavelets such as 
Mexican Hat, Morle, Gaussian wavelets and so on \cite{daubechies,ogden,meyer}. Usually they have infinite support and do not obey
exactly the Parseval's identity. Therefore such wavelets do not provide an
one-to-one reconstruction of a function from the its wavelet expansion coefficients. Due to these circumstances we do not use such basis sets in our work.  
\end{description}

Summing up the above, we can conclude that in order to provide good denoising of a signal the wavelets have to possess
good regularity and as many vanishing moments as possible. From other point 
of view, they have to be well localized in
space, which means that they must have a quite short support.
Unfortunately, these properties are interrelated. Thus, small support implies only few vanishing moments and poor
regularity. In addition, the orthogonality implies asymmetry of the basis functions, which in turn can lead to some numerical artefacts.
Since for each concrete task certain wavelet properties are more important than others, there are different wavelet families which are optimized 
for some of these properties.

For example, in the case of Daubechies wavelets we have a maximum number of vanishing moments and
maximal asymmetry with fixed
length of support, while the Symlet wavelet family has the "least asymmetry" and highest number of vanishing moments with a given
support width.

It was shown that it is possible to construct wavelet basis sets with the scaling function having
vanishing moments of non--zero order 
with respect to some shifting constant $c$.  Thus, for a  given number of vanishing moments $N_{v.m.}$ we have:
\begin{equation}
\int (r-c)^{n}\varphi(r)dr=0, \qquad 0<n<N_{v.m.}, \notag
\end{equation}
The Coifman wavelets are compactly supported wavelets which have the highest number of vanishing
moments for both $\varphi (r)$ and $\psi (r)$ with a given width of support.
This property is very useful for the treatment of functions with sharp peaks and slopes.
The larger the number of the scaling function  vanishing moments,
the better is the approximation for singular points of the function under study
\cite{daubechies}. Hence, by using such wavelets (e.g. Coifman) 
we can treat accurately
sharp peaks of such a function. On the other hand, these wavelets are rather
smooth to approximate well the function  within the ranges between these peaks.
The price for this extra feature is that the Coifman wavelets are longer than the Daubechies wavelets.
Their length of support is  equal to $3N_{v.m.}-1$ instead of $2N_{v.m}-1$.

Thus we can see that for orthogonal wavelets the  desirable properties are in contradiction to each other.
But fortunately, we can use different functions
for the decomposition and reconstruction. These biorthogonal bases have several
advantages compared with the orthogonal bases.
We can also benefit from the fact
that we can use base functions $\tilde{\varphi},\tilde{\psi}$
with a number of vanishing moments for decomposition,
whereas functions $\varphi,\psi$ with a good regularity
for reconstruction.
The former would separate any unpleasant stochastic oscillations
of TPCFs leaving this `noise' to the detail coefficients at higher
levels of resolution. The latter, on the other hand,
would produce a TPCF approximation as smooth as possible during reconstruction.
If, however, we would prefer to impose both conditions of a large
number of vanishing moments and regularity on an orthogonal basis,
we would have to pay with a support at least twice the size that of
the biorthogonal basis. Large supports, on the other hand, are known
to lead to a significant deterioration in the quality of the wavelet
approximation \cite{daubechies,donoho}.

In this work we will use biorthogonal bases from two biorthogonal
families:
Biorthogonal Spline Wavelets
whose  decomposition functions $\tilde{\psi}(r)$  are optimized for the number
of vanishing moments, but the reconstruction functions $\psi (r)$
are optimized in the sense of regularity; the
Reverse Biorthogonal Spline Wavelet whose
decomposition functions $\tilde {\psi} (r)$  are optimized to achieve maximal regularity with a given support width
and the  reconstruction functions $\psi (r)$  which are
constructed in order to gain a maximum number of vanishing moments.
In addition, these biorthogonal sets have the exact symmetry for all the basis functions.

\subsection{Wavelet Algorithm}

A typical way of building the wavelet approximation is as follows
\cite{donoho}. The coefficients obtained by FWT are sorted in
order of the decrease of their absolute values and then only some number
$L$ of the largest coefficients are kept by nullifying the rest of
the coefficients.
This is followed by application of the inverse transform (reconstruction).
Note that the truncation number $L$ depends on the required accuracy of
representation of the function in question. However,
this scheme is difficult to apply because of an
undesired intersection between different levels of resolution which
often arises. The latter leads to a much increased number of
coefficients required without any sensible improvement in the accuracy.
The quality of the resulting approximation is not particularly high
because the numerical boundary artifacts result in the so--called Gibbs effect,
i.e. false oscillations of the approximated function \cite{daubechies}.

Therefore, we will use instead a `smarter' strategy in which we employ
the following three remarkable circumstances:

\begin{enumerate}

\item For physical reasons the functions $\hat{g}_{ij}^{(2)}(\hat{r})$ vanish
at $\hat{r}\rightarrow 0$ (due to the excluded volume effect)
and at $\hat{r}\rightarrow \infty$ (due to a finite size of the molecule).

\item In terms of the rescaled radius, $\hat{r}\gtrsim 0.75$
the functions $\hat{g}_{ij}^{(2)}(\hat{r})$ have a rapid exponential
(or even a faster stretched exponential) decay.

\item From physical grounds it is also well known that
$\hat{g}_{ij}^{(2)}(\hat{r})$ is differentiable function of a high order
for large $\hat{r}$.

\item The multi--resolution nature of the wavelet analysis allows us to treat
each level of the wavelet decomposition separately.

\end{enumerate}

We have developed an advanced scheme of the wavelet approximation
which, first of all, takes into account the peculiarities of TPCF.
From another side it relies on the strategy of a `level-by-level' thresholding,
which has been independently proposed by several authors
\cite{donoho,ogden}.

By taking into account the asymptotic behavior of TPCFs, we can use
the zero boundary conditions while doing the wavelet decomposition.
Considering the values of $\hat{g}(\hat{r}\rightarrow 0)$ as zero,
we can also nullify all wavelet coefficients corresponding to the range
$[0,0.05]$.
Strictly speaking, the upper bound for this cut--off is given by
$\hat{r}_l \equiv d/\sqrt{D_{ij}}$ and it depends on the system size and
parameters, but the value of $0.05$ is well below this bound for all the data
considered in this paper.
As we have decomposition functions with a sufficient number of vanishing
moments, we can nullify all detail coefficients at all levels of
resolution which correspond to the range of rescaled radius $\hat{r}\in
[0.75,...,\infty)$ in order to extract the trends of our TPCF with a
`maximal smoothness' \cite{daubechies,belkyn}.
The value for this lower bound $\hat{r}_r$ has the meaning of the rescaled
radius after which TPCF has a fairly smooth decaying behavior.
For other regions of
$\hat{r}$ we extract the \emph{highest} detail coefficients in
\emph{each} levels of resolution \emph{separately}.

Summing up all of the above, we propose the following scheme for the TPCF
wavelet approximation:
\begin{enumerate}
\item We perform FWT with zero boundary conditions at the largest scale $M$
satisfying the condition
$\sum_{b}|d_{M,b}|\leq \epsilon \sum_{b}|a_{M,b}|$ (where a good
choice for $\epsilon$ is $0.05$),  then all
further $d$-coefficients can be neglected.
\item All the coefficients corresponding to the range $r\in [0,0.05]$
(for both the approximation and the detail) are also nullified.
\item We save \emph{all} the approximation coefficients which
remain non--vanishing in the previous steps.
\item All the \emph{detail} coefficients corresponding to $r\in [0.75,...,\infty)$
are nullified.
\item In each level of decomposition we leave the maximal detail coefficients
corresponding to the function extrema, while neglecting the rest of the
coefficients.
\item We perform the conventional inverse FWT but only for the
non--zero coefficients remaining from the previous steps.
\item To suppress the Gibbs effect at the left boundary, the approximated
TPCF is set equal to zero up to $r_{cross}$, where $r_{cross}$ is the
rightmost nontrivial zero point of the approximated TPCF,
i.e., $g_{app}(r_{cross})=0$.
\end{enumerate}
As a result, we have a fast scheme of calculations and a compact
approximation for the correlation functions.

\begin{figure*}
\resizebox{14cm}{!}{
\includegraphics{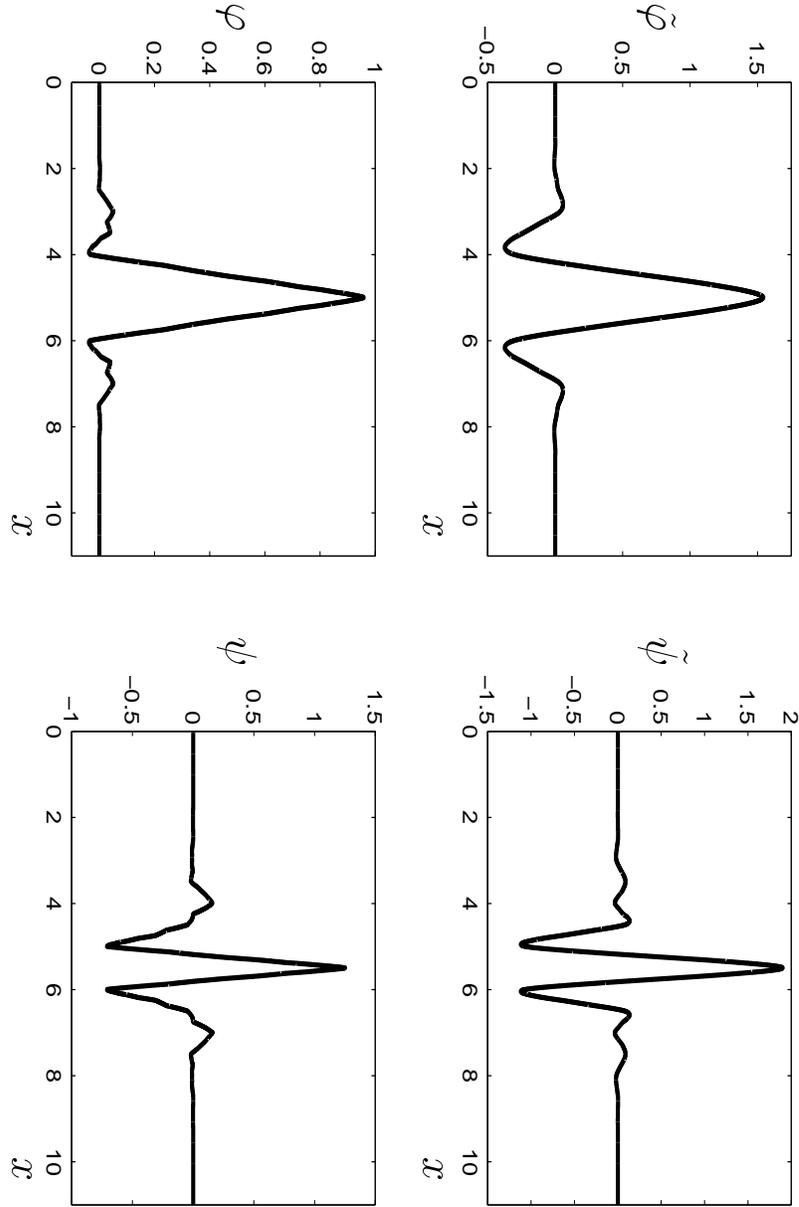}
}
\caption{\label{fig:1}
Reverse Biorthogonal Spline Wavelets 5-5. The abscissa is the real
numbers axis ($x \in \mbox{{\msbmtex R}}$).
At the top are the decomposition scaling function $\tilde{\varphi}$
and the wavelet function $\tilde{\psi}$ and at the bottom are the
corresponding reconstruction functions $\varphi$ and $\psi$.
Here and in all other figures the axes are depicted in dimensionless units.
}
\end{figure*}

Concerning the choice of the wavelet basis set, we note that
to realize FWT there are many suitable sets such as Daubechies,
Coifman, Symlets, biorthogonal wavelets, and so on
\cite{daubechies,sweldens}.
We have tested various basis sets, but our detailed
study presented below indicates that the reverse biorthogonal basis
(RB5-5) is the best of them for treatment of TPCF for the systems under
study. Here we shall follow the
Daubechies' notation for this family: the first index --- $N_{d}=5$ for the
decomposition functions, the second index --- $N_{r}=5$ for the
reconstruction ones. These indices reflect the number of vanishing
moments of $\tilde{\psi}$, namely: $N_{v.m.}=N_{r}-1$, the regularity
value of $\psi$, namely $\rho=N_{r}-1$, as well as the length of support $l$
for the pairs $\{\tilde{\varphi},\tilde{\psi}\}$:
$l_{d}=2*N_{d}+1$, and for the pairs $\{\varphi,\psi\}$: $l_{r}=2*N_{r}+1$.
Fig.~\ref{fig:1} depicts the functions from the RB5-5 basis set.

\section{Results}

\begin{figure*}
\resizebox{14cm}{!}{
\includegraphics{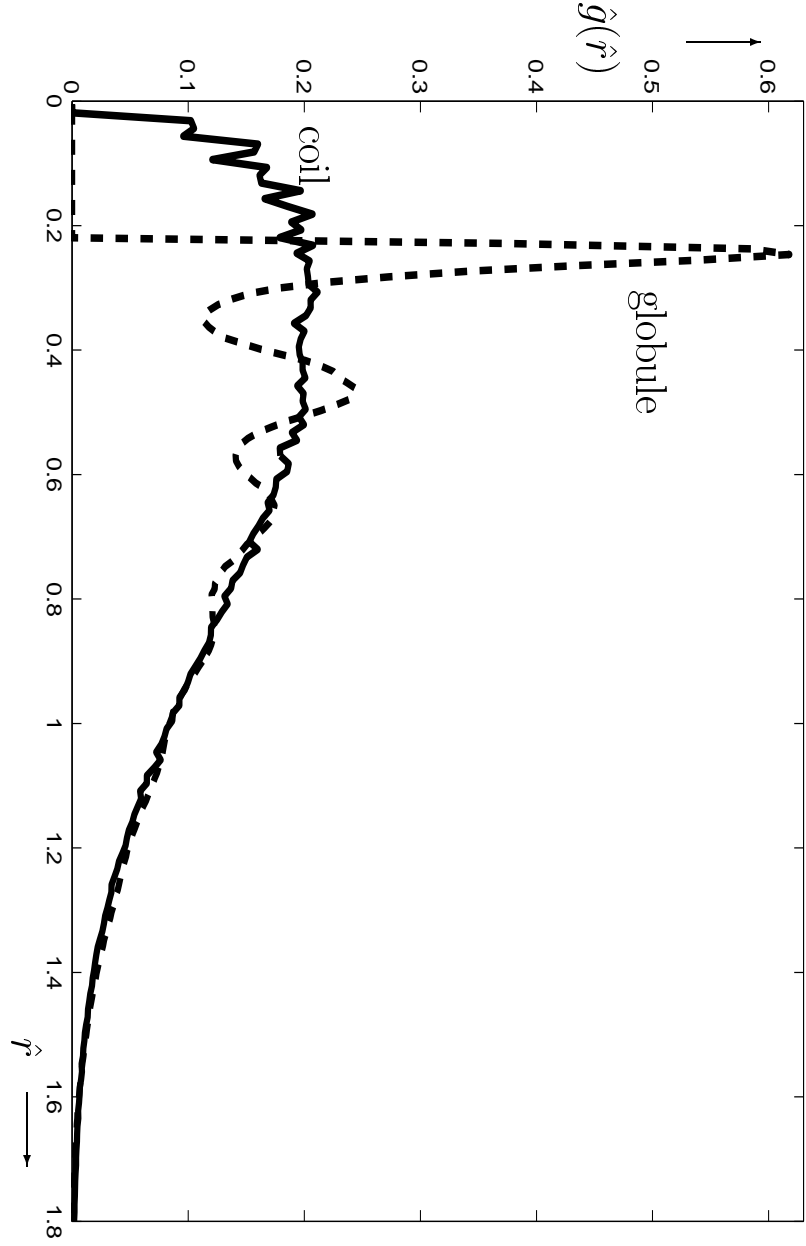}
}
\caption{\label{fig:2}
Rescaled correlation function of the homopolymers with the
degree of polymerization $K=200$. The
solid curve corresponds to the end--end correlations of
of an open homopolymer in the coil state.
The dashed curve corresponds to the globule of a ring homopolymer
with $K=200$ and for $|i-j|=100$.
}
\end{figure*}

To illustrate the usefulness of our scheme we have investigated
the two--point correlation functions of ring, linear and star homopolymers in
the coil state, as well as of the globular state of a ring
homopolymer since the connectivity is not as important for the latter
state.
The data for $\hat{g}_{ij}^{(2)}(\hat{r})$ has been obtained by Monte Carlo
simulations discussed in our previous study \cite{IntraCor}.
Fig.~\ref{fig:2} depicts typical behavior
of $\hat{g}_{ij}^{(2)}(\hat{r})$ for an open homopolymer coil and a
ring homopolymer globule. As one can see, in the liquid globular state the
TPCF has several peaks of
increasing width and decreasing height located at approximately
$n\,d/\sqrt{D_{ij}}$
($n=1,2,\ldots$).
On the other hand, the
TPCF of a coil exhibits a smoother radial dependence, but suffers from a
significant statistical noise.

\begin{figure*}
\resizebox{14cm}{!}{
\includegraphics{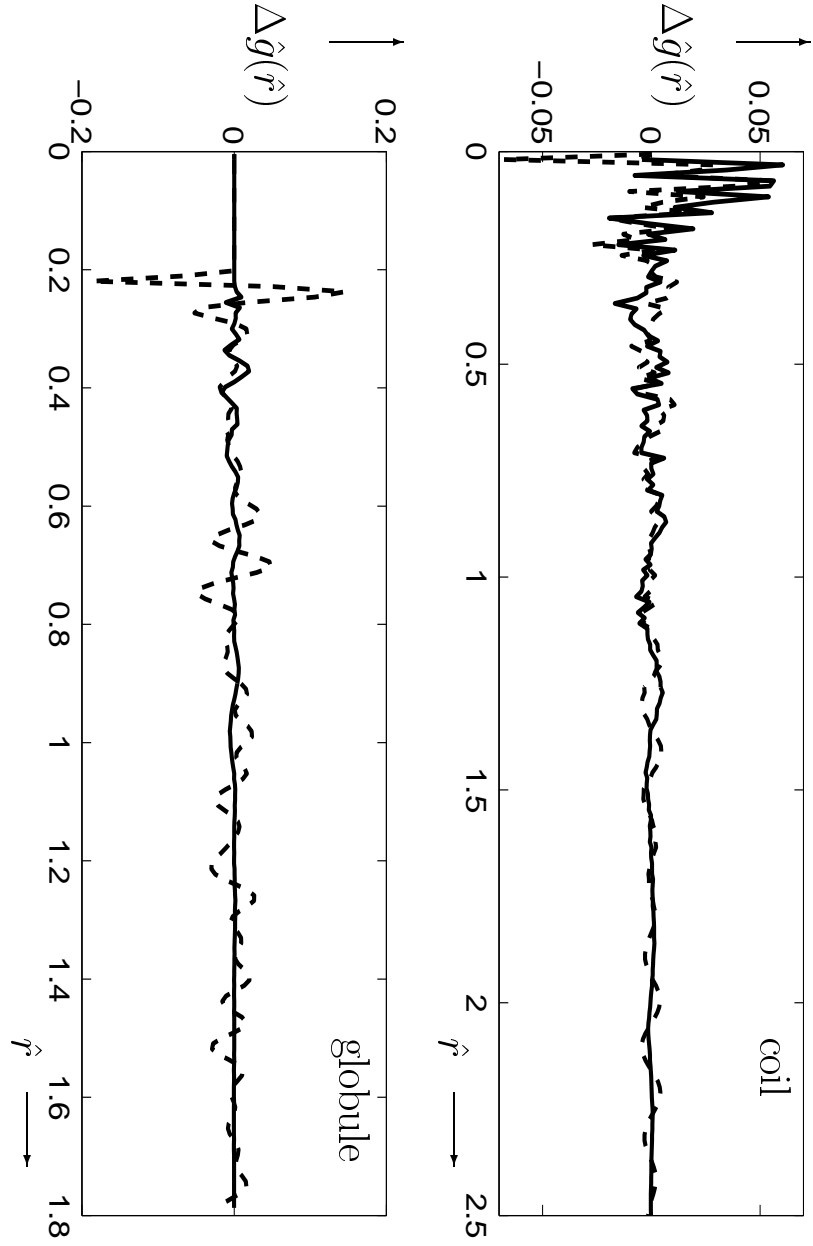}
}
\caption{\label{fig:3}
The difference $\Delta \hat{g}(\hat{r})$ between the TPCF obtained from
simulations and their approximations by wavelets (solid curve) and
cosines (dashed curve). The upper part of the figure corresponds to
the coil of an open homopolymer, while the lower one to the globule of a ring
homopolymer of the same lengths as in Fig. \ref{fig:2}.
}
\end{figure*}

The correlation functions obtained from such data are then approximated
by the above described wavelet procedure. Fig.~\ref{fig:3} shows the difference
$\Delta \hat{g}(\hat{r})=\hat{g}(\hat{r})-\hat{g}_{app}(\hat{r})$
of the TPCFs obtained by simulations and their
approximations by wavelets (solid curves) and cosines (dashed
curves) with the same number of terms $L$. 
One can clearly see that the wavelet treatment provides
a much better approximation than the cosines Fourier treatment.
For the coil ($L=20$), at small radial separations both treatments do show
deviations from the simulation data, but these only reflect the
limitations of sampling statistics of TPCF as the function should
really be very smooth and obey the des Cloizeaux equation.
However, while the wavelet treatment gives an essentially vanishing
$\Delta \hat{g}$ for larger $\hat{r}$, the Fourier treatment continues
to yield parasitic oscillations at all separations.
For the globule ($L=25$),
which had a much better quality of data due to a smaller
entropy of the globule, the wavelet treatment gives an essentially
vanishing $\Delta \hat{g}$ everywhere, whereas the Fourier method
works very poorly in the whole range with strong oscillations present even
for the largest of separations.

\begin{figure*}
\resizebox{14cm}{!}{
\includegraphics{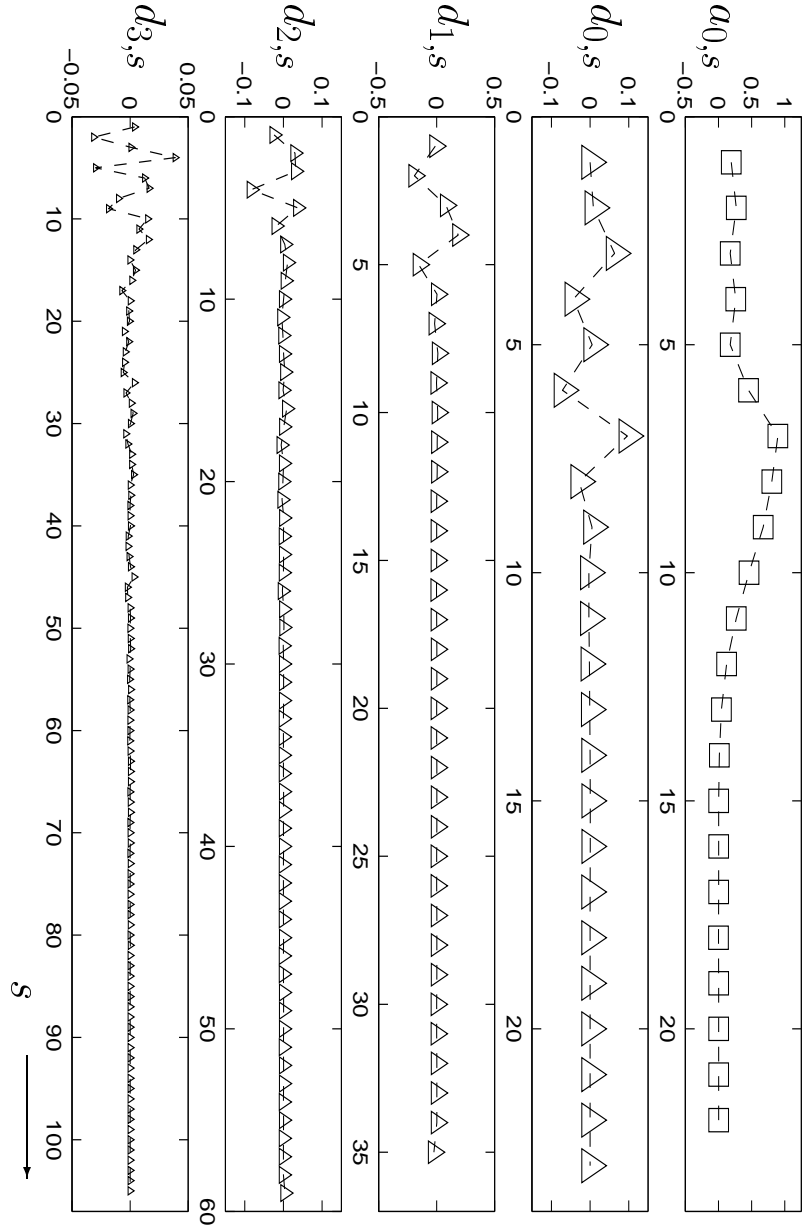}
}
\caption{\label{fig:4}
Four different levels of the wavelet decomposition of the end--end
TPCF of an open homopolymer with $K=200$.
\\
At the top there are approximating coefficients $\{a\}$ at the level
$j_{0}=0$.
The detail coefficients $\{d\}$ are presented in the ascending
order in the level of
the resolution $j$ vs the shift parameter $s$.
}
\end{figure*}

In Fig.~\ref{fig:4} we present four different levels of the wavelet
decomposition of TPCF of an open coil.
We can see that the smooth part of this function can be well
represented by the approximation coefficients. Conversely, the
unpleasant oscillations are concentrated in the detail
coefficients.

\begin{figure*}
\resizebox{14cm}{!}{
\includegraphics{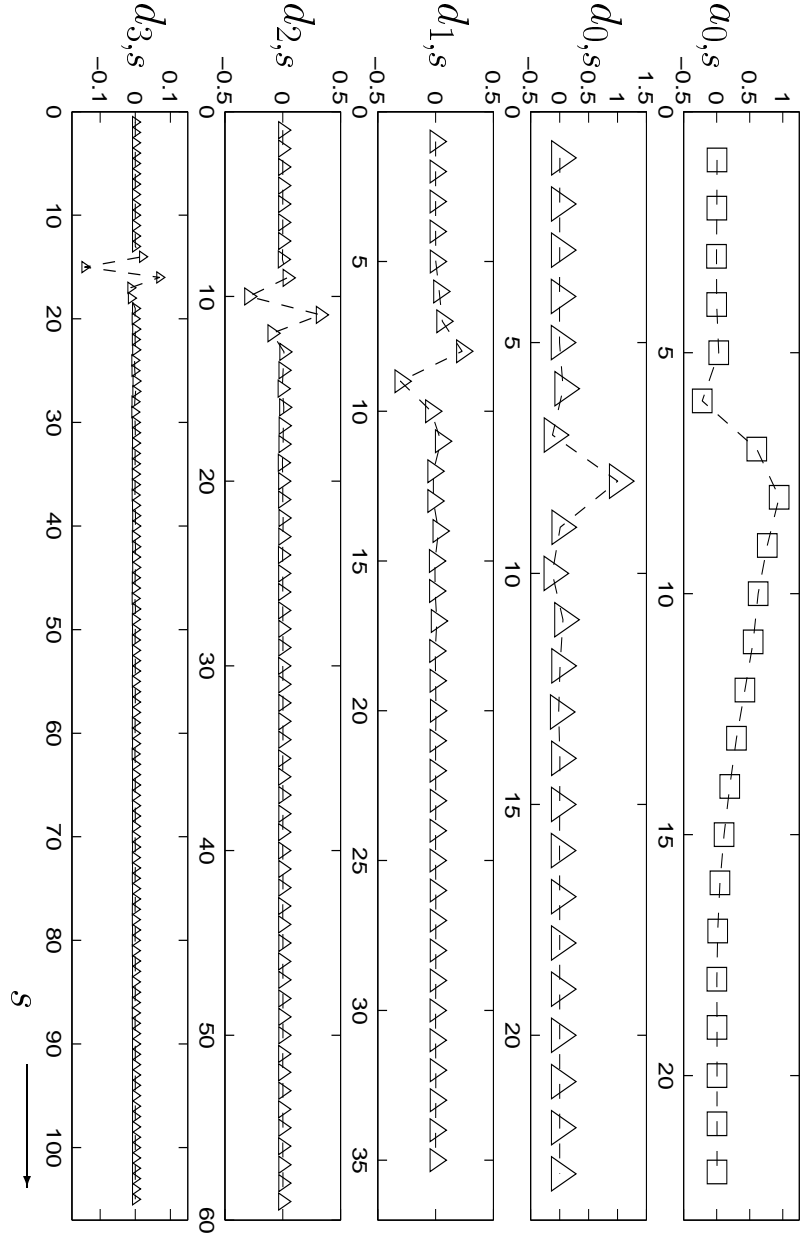}
}
\caption{\label{fig:5}
Four different levels of the wavelet decomposition of TPCF of the homopolymer
globule of the same lengths and for the same chain indices as in Fig. \ref{fig:2}.
At the top there are approximating coefficients $\{a\}$ at the level
$j_{0}=0$.
The detail coefficients $\{d\}$ are presented in
the ascending order in the level of
the resolution $j$ vs the shift parameter $s$.
}
\end{figure*}

In Fig.~\ref{fig:5} we likewise present four different levels of the wavelet
decomposition of TPCF for the globule of a ring homopolymer.
We can see that the smooth part of this function can be mainly
represented by the approximation coefficients.
But there is also an important information in the
detail coefficients, which mainly represent the sharp peaks of the function.
Therefore, our `smart' level--by--level technique allows us to
effectively suppress noise in case of the coil and to
prevent us from `oversmoothing' of physical oscillations
in case of the globule.

\begin{figure*}
\resizebox{14cm}{!}{
\includegraphics{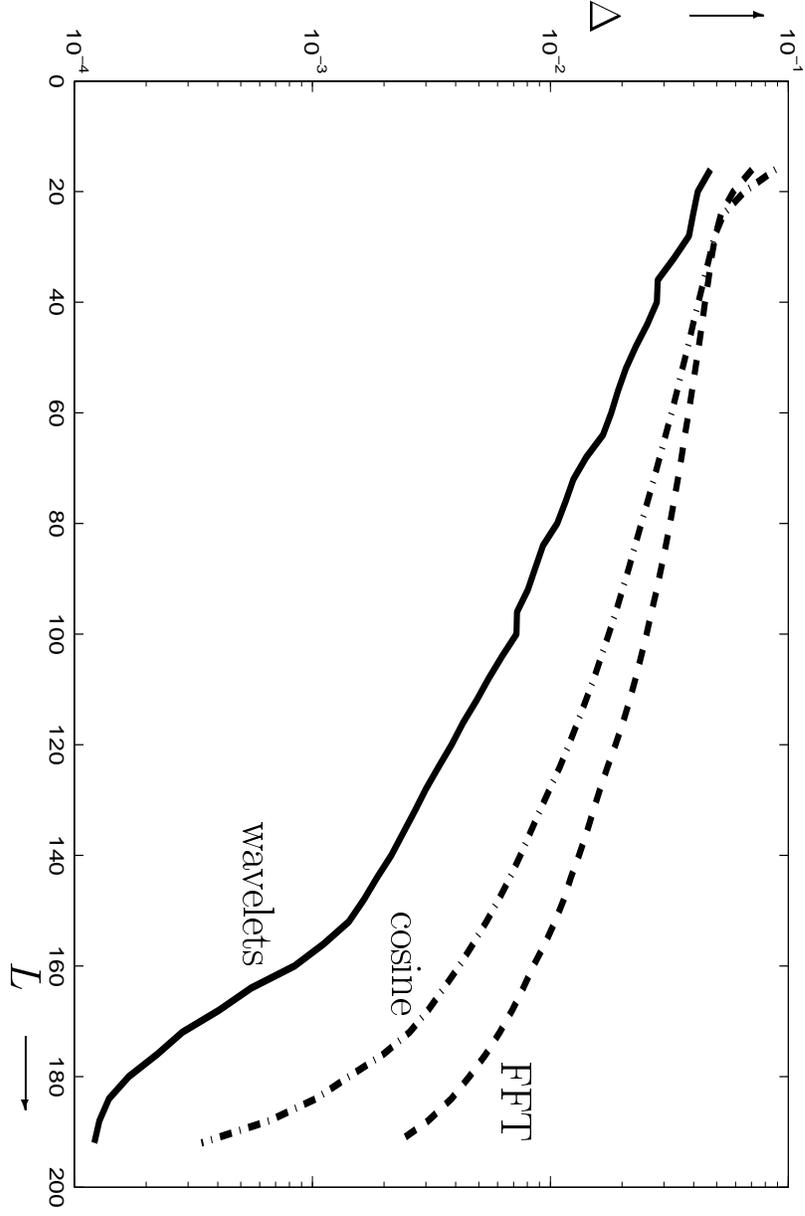}
}
\caption{\label{fig:6}
Square root error $\Delta$ between the rescaled end--end TPCF
of a homopolymer ring in the coil state with the
degree of polymerization $K=200$ and
its approximations. The curves correspond to the wavelet approximation
(solid line), FFT approximation (dashed line) and cosine
approximation (dash--dotted line). The X-axis is the total number $L$ of
the approximation coefficients used.
}
\end{figure*}

\begin{figure*}
\resizebox{14cm}{!}{
\includegraphics{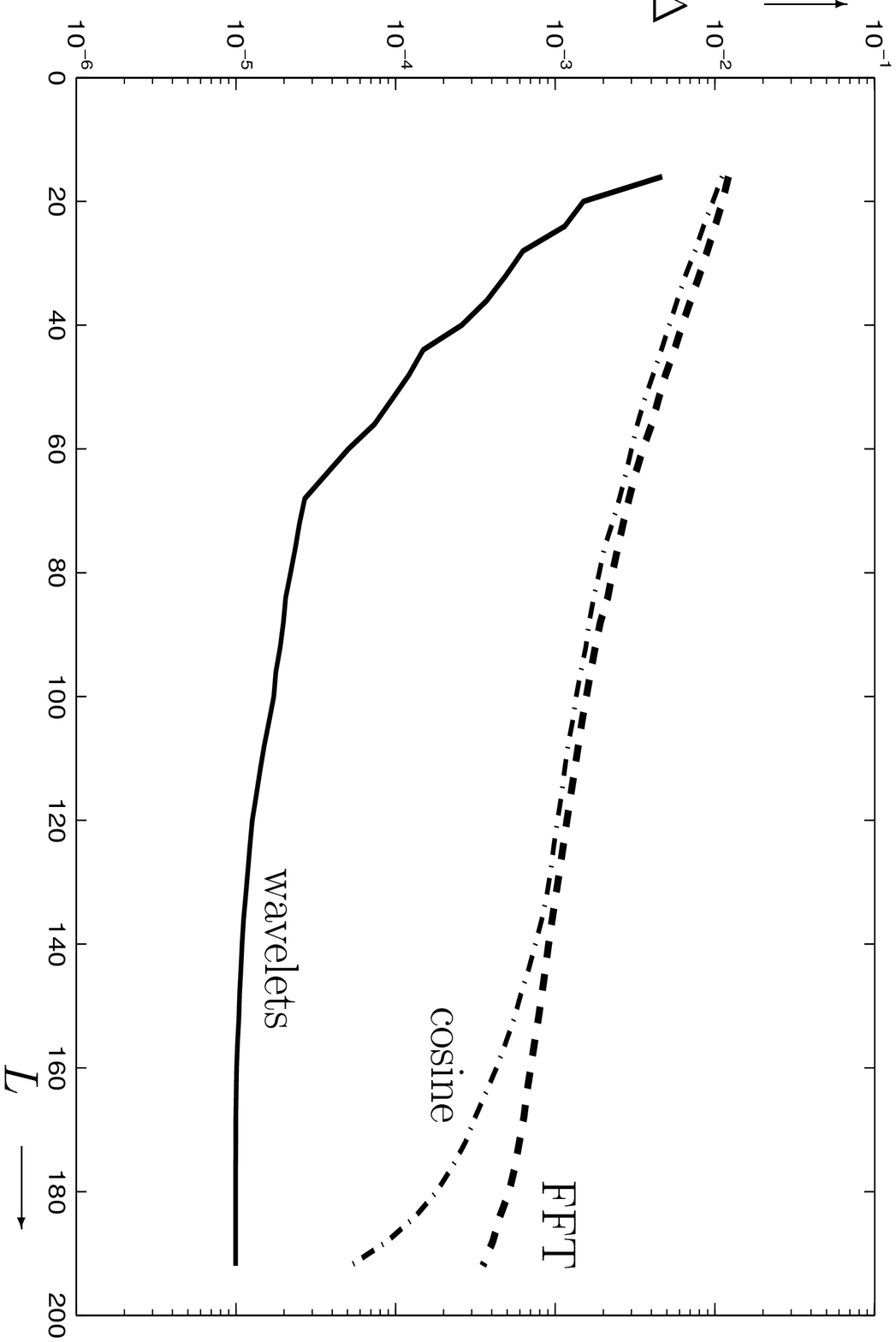}
}
\caption{\label{fig:7}
Square root error $\Delta$ between the rescaled
TPCF of a homopolymer ring with $K=200$ and $|i-j|=100$
in the globular state and
its approximations.
\\
The curves correspond to the wavelet approximation (solid
line), FFT approximation (dashed line) and cosine
approximation (dash--dotted line). The X-axis is the number $L$ of
approximation coefficients used.
}
\end{figure*}

We have also calculated the mean square norm of the inaccuracy $\Delta$,
which characterizes the quality of the approximation:
$\Delta\equiv\sqrt{\sum_{i=1}^{n}(\hat{g}(\hat{r}_{i})
-\hat{g}_{app}(\hat{r}_{i})})^{2}$, where
$\hat{r}_{i}=i\cdot\delta \hat{r}$ are the grid points,
$\hat{g}(\hat{r}_{i})$ is the `true' correlation
function from Monte Carlo data, and $\hat{g}_{app}(\hat{r}_{i})$
is the approximated one.
Figs.~\ref{fig:6} and \ref{fig:7} depict the dependences of the norm
$\Delta$ on the number $L$ of the approximating coefficients for the
coil and globule states respectively.
In what was mentioned above we have used
the `RB5-5' basis set \cite{daubechies}. Here, for comparison we also
depict these dependencies for the Cosine and FFT approximation,
which are widely used in applications \cite{marpl}. We can see
that for a reasonable number of coefficients the wavelet
approximations gives us a remarkably better accuracy than the
conventional methods.

For the approximated TPCF we have also evaluated the scaling exponents
for the coil state. In this case we can compare these results with the
rather accurate theoretical values obtained from the Borel resumed
renormalisation group calculations
\cite{CloizeauxBook,Schafer,Guida,Caracciolo}.
As in Ref. \cite{IntraCor}
the fitting has been done via the the nonlinear
least--squares (NLLS) Marquardt--Levenberg method
\cite{NumerRecip} by means of the {\tt fit} function in the {\tt
gnuplot} software.
{\tt Fit} reports parameter error
estimates which are obtained from the variance--covariance matrix
after the final iteration. By convention, these estimates are
called `standard errors' and they have been reported in
Tab.~\ref{tab:1}, which
contains the results for open and ring homopolymers.
Here we have used the wavelet approximation with $L=20$ coefficients.
In this table we also include the results which are obtained from the
fitting of the untreated functions as in Ref. \cite{IntraCor}.
The notations in the first column follow the
des Cloizeaux convention: \textbf{0}
--- end--end monomers, \textbf{1} --- end--middle, \textbf{1}' ---
end--three quarters, \textbf{1}'' --- end--one quarter, and \textbf{2}
--- one quarter--three quarters of the chain respectively. Here and
below reported errors are those from the fitting procedure only and
do not necessarily account for statistical and other simulation errors.

We can see that the wavelet approximated functions
agree with the most recent theoretical values much better.
Note also that some of the theoretical values
in this table have been updated
thanks to the more accurate values from Refs. \cite{Schafer,Guida,Caracciolo}
as compared to those which we have used in Ref. \cite{IntraCor}.
Moreover, we do not even need to freeze
$\delta$ at the theoretical value in order
to extract a more accurate estimate for $\theta$ as we had to do previously.
These improvements in the results of our fitting are not surprising
given that, as we have mentioned earlier, the coefficients cut--off leads to
an effective noise suppressing.

The least--squares fitting of the data $\{{\bf x},{\bf y}\}$
with a model function
$y(x_i; {\bf a})$, which
depends on the fitting parameters ${\bf a}$ in a nonlinear fashion, in
the multi--variate case is a complex problem \cite{NumerRecip}
akin to that of finding the global minimum of the merit function
$\chi^2({\bf a})
=\sum_{i=1}^{N}\sigma_i^{-2}(y_i -y(x_i; {\bf a}))^2$ with
respect to $N$ parameters ${\bf a}$, where
$\sigma_i$ is the standard deviation (error) of the i-th data point.
If the data is fairly noisy, the problem of finding the global
minimum of $\chi^2$ becomes complicated as there are many low--lying
local minima of this function and its constant value surfaces have
a complicated topology.
The Marquardt--Levenberg method is one of the most popular fitting algorithms
which is an efficient hybrid of the inverse--Hessian (variable metric) and
the steepest descent (conjugated gradients) minimization
algorithms for $\chi^2$  \cite{NumerRecip}.
Practically, the iterations need to be stopped
after the values of $\chi^2$ change less than the specified precision
and, clearly, the resulting fitted values ${\bf a}_{fit}$ may depend
on the choice of the initial values ${\bf a}_{0}$ if there are many
local minima present, as well as on the weights ${\bf \sigma}$. It is not
uncommon to find the parameters wandering around near the global minimum
in a flat valley of complicated topology if the input data was fairly noisy
\cite{NumerRecip}. The wavelet treatment renders the initial
poorly--defined fitting problem into a well--defined one
(which becomes essentially independent of the initial parameters choice)
by removing the high--pitch statistical
noise from the data, and thus by simplifying the topology of the constant
$\chi^2$ surfaces and getting rid of its many artificial local
minima. At the same time, the variances (squared standard errors)
of the fitted  parameters and the co-variances between them become
smaller than for the untreated data as we are now guaranteed to
have found the true $\chi^2$ global minimum.

Tab.~\ref{tab:2} lists similar exponents
for star homopolymers with the number of arms $f=12$
in a good solvent. Here we have
used the wavelet approximation with $L=30$ coefficients.
We then have compared the results with those from the analytical
renormalisation group calculations in the so--called `cone' approximation
for $\theta$ from Ref. \cite{Ferber}.
This comparison has not been previously made in Ref. \cite{IntraCor}
or elsewhere so far, to the best of our knowledge.
The agreement between the Monte Carlo and theoretical values
seems quite reasonable despite the relatively short length of the
arms and the limitations of the `cone' approximation. The latter produces
the contact exponents only
dependent on the functionalities $f_i$, $f_j$ of the two monomers in question
and not on any other parameters of the star, namely
\begin{equation}
\theta_{f_i,f_j}\simeq \frac{5}{36}\frac{1}{\sqrt{2}-1}\left(
(f_i+f_j)^{3/2}-f_i^{/3/2}-f_j^{3/2}\right).
\end{equation}
As one can see, the quality of the wavelet approximation is rather good for the
combined scheme, while the number of approximating coefficients is
quite small.

In general, the accuracy of the wavelet approximations with a fixed
number of reconstruction coefficients depends on the chosen basis set.
So far, no exact `recipe' was given on which base we have to
use in a concrete case. Thus, we have checked our assumption about one of
the dual bases `RB5-5' by an additional study. As we are especially interested
in the quality of the scaling exponent calculations we have evaluated the
scaling exponent for an open homopolymer coil with the use of different
bases. These results are presented in Tab.~\ref{tab:3}. We have used the
wavelet approximations of the end--end TPCFs with the same number of
coefficients. We chose for this comparison typical representatives of
the main wavelet families:
Coifman 2, Daubechies 4, Symlet 4, Biorthogonal 5-5 and
Discrete Meyer wavelets \cite{daubechies}.

We can see that the `Reverse Biorthogonal 5-5'  basis set does
the approximation better than the other bases.
On the other hand, other bases, apart from the Discrete Meyer's, also reveal
good fitting results compared to the untreated TPCF.
This means that our current scheme is just one of possible successful
choices of the basis set.
The situation with the
Discrete Meyer is easily explained by a too large support
length of this basis $l=60$ as compared with
$l=11$ for `RB5-5', which is known to lead to strong over--smoothing
\cite{daubechies}.

\section{Conclusion}

Our present study indicates that the discrete wavelets
is a suitable and powerful instrument for approximating
the intra--chain two--point correlation functions (TPCF)
of different homopolymers in dilute solutions.
The wavelet technique allows us to extract the
scaling properties from fairly noisy data more accurately
and reliably than it can be done by the direct fitting.
The wavelet treatment removes the high--pitch stochastic fluctuations
(part of `statistical noise') in the data thereby producing a somewhat
`coarse--grained' approximation of the data.
This renders the ill-defined multi--variate nonlinear fitting procedure
of the untreated data into a well--defined uniquely convergent
fitting procedure after the wavelet treatment of the data.
Naturally, this also reduces the standard deviations of the fitted parameters.
However, the wavelet treatment does not over--smooth the data by
retaining the genuine oscillations as we clearly see in the case of the
globule, nor does it produce any of the unpleasant artefacts of the
truncated Fourier approximation.

We can see that the dual basis set performs particularly well for approximating
TPCFs. This is related to the basis properties, namely, that the decomposition
functions have a maximal number of vanishing moments with a finite
support, whereas the reconstruction functions are as regular as possible with
a given length of support.
Moreover, the proposed scheme is rather flexible as it is
based on the conventional FWT algorithm. One can
choose the basis set and adjust the number of the coefficients easily
for a particular problem.
From the results in Tab.~\ref{tab:3} we can also conclude that by
using almost any reasonable basis it is possible to obtain
an improvement in the fitting procedure.

It should be emphasized that the wavelet scheme is rather universal.
The scheme of the wavelet approximation proposed here allows us to
represent the correlation functions with a small number of approximating
coefficients not only for the coil but also for globular state of the
homopolymers.
For instance, our procedure yields
the relative accuracy of the approximation of order
$\frac{1}{n}\Delta\sim 0.5\cdot10^{-3}$. Such accurate knowledge of
TPCFs can be used for an input to the self--consistent calculations of
the inter--chain distribution functions in the framework of the density
functional methods \cite{yet98,yet03,sweat} and others.

Due to a compact parameterization and a high accuracy of the
approximation by wavelets, we hope that the wavelets can be applied not
only for approximating the inter--chain distribution functions
of polymers, but also in order to calculate these functions by
the integral equations theory of polymers \cite{monson}. The success of
the recent applications of wavelets
to the theory of molecular solutes \cite{chuevfedorov1}
has indicated that the method is capable of
calculating the thermodynamic characteristics of solvation rather accurately.

We believe that further progress in this direction can also be of
importance for the novel theories for calculating the intra--chain TPCFs
of polymers directly from a force field. Some of us are presently
working on the Super--Gaussian Self--Consistent (SGSC) theory for
a single macromolecule with any two--body Hamiltonian, in which
a set of integro--differential equations is derived for $\hat{g}_{ij}(\hat{r})$
as well as for $D_{ij}$. In order to reduce the computational expenses
in such calculations, having a compact and multi--resolution accurate
representation for $\hat{g}_{ij}(\hat{r})$ is essential.

Finally, due to the very general nature of the wavelet theory, we hope that
wavelets can find other numerous applications for describing spatial and
temporal dependences of various observables in a number of fields
of soft condensed matter theory which they have not hereto
beneficially influenced.

\begin{acknowledgments}
We would like to thank Professor L.~Sch\"{a}fer,
Dr C. von Ferber, Dr H.-J. Flad, Dr H. Luo
and Professor D. Kolb  for interesting discussions.
We acknowledge the support of the Centre for Synthesis and
Chemical Biology, and some of us (M.F. and G.C.) also
acknowledge the support of the Russian Foundation for Basic Research.

\end{acknowledgments}

\newpage


\begin{table*}
\caption{
\label{tab:1}
Comparison of the exponents $\delta$ and $\theta$
between the results from the direct fitting of the Monte Carlo data with
those from their wavelet approximations (subscript $ww$) and the theoretical
results (subscript $theor$) for open and ring homopolymer coils
with the degree of polymerization $K=200$. Fitted values
have been obtained by a four--parametric
fit via Eq. (\ref{FitA}).
}

\begin{tabular}{|c|c|c|c|c|c|c|}
\hline
&
 $\delta$&
$\delta_{ww}$&
$\delta_{theor}$&
$\theta$&
$\theta_{ww}$&
$\theta_{theor}$\tabularnewline
\hline
\hline
\textbf{0}&
$2.11\pm0.07$&
$2.36\pm0.02$&
$2.428\pm0.001$&
$0.36\pm0.02$&
$0.276\pm0.005$&
$0.271\pm0.002$\tabularnewline
\hline
\textbf{1}&
$2.23\pm0.04$&
$2.36\pm0.02$&
...&
$0.56\pm0.01$&
$0.51\pm0.01$&
$\sim0.46$\tabularnewline
\hline
\textbf{1'}&
$2.42\pm0.04$&
$2.39\pm0.02$&
...&
$0.45\pm0.01$&
$0.462\pm0.003$&
$0.459\pm0.003$\tabularnewline
\hline
\textbf{1"}&
$2.04\pm0.08$&
$2.39\pm0.02$&
...&
$0.68\pm0.03$&
$0.52\pm0.02$&
$\sim0.46$\tabularnewline
\hline
\textbf{2}&
$2.39\pm0.07$&
$2.40\pm0.02$&
...&
$0.81\pm0.02$&
$0.80\pm0.01$&
$0.80\pm0.01$\tabularnewline
\hline
\textbf{Ring}&
$2.46\pm0.07$&
$2.40\pm0.02$&
...&
$0.79\pm0.006$&
$0.815\pm0.005$&
$0.80\pm0.01$\tabularnewline
\hline
\end{tabular}
\end{table*}

\begin{table*}
\caption{
\label{tab:2}
Values of the exponents $\delta$ and $\theta$ for star homopolymers with $f=12$
arms and $(N-1)/f=50$ arm length in a good solvent.
The following notations for the monomer pairs have been adopted:
$a:n,b:m$, where $a,b$ number arms and $n,m$ number monomers within arms
and $0$ refers to the core monomer. Other notations are the same as in Tab.
\ref{tab:1}.
}
\begin{tabular}{|c|c|c|c|c|c|}
\hline
&
$\delta$&
$\delta_{ww}$&
$\theta$&
$\theta_{ww}$&
$\theta_{theor}$
\tabularnewline
\hline
\hline
\textbf{0, a:m=25}&
$2.67\pm0.03$&
$2.67\pm0.002$&
$3.036\pm0.04$&
$3.024\pm0.02$&
$2.677$\tabularnewline
\hline
\textbf{0, a:m=50}&
$2.56\pm0.03$&
$2.62\pm0.01$&
$0.59\pm0.02$&
$1.51\pm0.01$&
$1.442$\tabularnewline
\hline
\textbf{a:n=25, a:m=50}&
$2.40\pm0.06$&
$2.48\pm0.004$&
$0.55\pm0.01$&
$0.50\pm0.003$&
$0.458 $\tabularnewline
\hline
\textbf{a:n=50, b:m=50}&
$2.30\pm0.03$&
$2.18\pm0.01$&
$0.23\pm0.01$&
$0.26\pm0.007$&
$0.277$\tabularnewline
\hline
\end{tabular}
\end{table*}

\begin{table*}
\caption{\label{tab:3}
Comparison of the exponents $\delta$ and $\theta$
between the theoretical results,
different wavelet approximations (with names of the wavelet bases are
given in the first column), and
the untreated results from the direct fitting of Monte Carlo data
for the end--end TPCFs of an open homopolymer coil
with the degree of polymerization $K=200$.
}
\begin{tabular}{|c|c|c|}
\hline
                     & $\delta$        & $\theta$        \tabularnewline
\hline
\textbf{Theoretical} & $2.428\pm0.001$ & $0.271\pm0.002$ \tabularnewline
\hline
\hline
\textbf{`RB5-5'}     & $2.36\pm0.02$   & $0.276\pm0.005$ \tabularnewline
\hline
\textbf{`BR5-5'}     & $2.36\pm0.02$   & $0.31\pm0.01$   \tabularnewline
\hline
\textbf{Daubechies 4}& $2.35\pm0.02$   & $0.29\pm0.02$   \tabularnewline
\hline
\textbf{Symlet 4}& $2.34\pm0.02$   & $0.28\pm0.02$   \tabularnewline
\hline
\textbf{Coifman 2}   & $2.36\pm0.02$   & $0.285\pm0.005$ \tabularnewline
\hline
\textbf{Discr. Meyer}& $2.1\pm0.01$    & $0.4\pm0.01$    \tabularnewline
\hline
\hline
\textbf{Untreated}   & $2.11\pm0.07$   & $0.36\pm0.02$   \tabularnewline
\hline

\end{tabular}
\end{table*}

%
%
%

\newpage


\end{document}